\documentstyle[prd,aps]{revtex}

\def\pars{{\cal P}}
\def\drxn{{\bf n}}
\def\like{{\cal L}}
\def\drxn{{\bf n}}
\def\svec{{\bf s}}
\def\mod{\mathop{\rm mod}}

\begin{document}
\draft

\title{Bayesian Analysis of the Polarization of Distant Radio Sources:\\
Limits on Cosmological Birefringence}

\author{Thomas J. Loredo, Eanna E. Flanagan, and Ira M. Wasserman}
\address{Department of Astronomy,
Space Sciences Building,
Cornell University,
Ithaca, New York 14853}

\date{22 June 1997}

\maketitle

\begin{abstract}
A recent study of the rotation of the plane
of polarization of light from 160 cosmological sources claims to
find significant evidence for cosmological anisotropy.  We point
out methodological weaknesses of that study, and reanalyze
the same data using Bayesian methods that overcome these problems.
We find that the data always favor isotropic models for the
distribution of observed polarizations over counterparts that
have a cosmological anisotropy of the type advocated in the
earlier study.  Although anisotropic models are not completely
ruled out, the data put strong lower limits on the length
scale $\lambda$ (in units of the Hubble length) 
associated with the anisotropy; the lower limits of
95\% credible regions for $\lambda$ lie between 0.43 and 0.62
in all anisotropic models we studied, values several times
larger than the best-fit value of $\lambda \approx 0.1$ 
found in the earlier study.
The length scale is not constrained from above.
The vast majority of sources in the data are at distances closer
than 0.4 Hubble lengths (corresponding to a redshift of $\approx$0.8);
the results are thus consistent with there being no significant
anisotropy on the length scale probed by these data. 
\end{abstract}

\pacs{98.80.Es, 41.20.Jb, 02.50.Ph}

Nodland and Ralston \cite{NR97}\ recently analyzed the distribution of
the intensity-averaged polarization angles of 160 galaxies and claim to
find significant evidence for
cosmological birefringence---a systematic tendency for the plane of
polarization of light to rotate in a manner dependent on the direction
of travel and proportional to the distance traveled.  The magnitude
of the claimed effect is large; the predicted cosmological
polarization angle rotation for a source at a Hubble distance can
be as large as several radians, depending on the direction to the source.
Such a discovery would be of immense importance if true.  However, recent 
analyses of other data better suited to testing the
Nodland-Ralston hypothesis find no
evidence for cosmological birefringence and convincingly
demonstrate that, if present, it must have a
magnitude at least 30 to 100 times smaller than that claimed
by Nodland and Ralston \cite{Leahy97,WPC97}.

These analyses leave open the issue of accounting for the effect 
Nodland and Ralston
claim to detect in the data set they analyzed.  We argue here that the
effect is not present and that
methodological weaknesses corrupted their study (some similar
points have been made in \cite{EB97,CF97}), and we
demonstrate this by reanalyzing the same data in the framework of
Bayesian inference where the weaknesses we identify are automatically
dealt with and overcome.  Our analysis finds no evidence for cosmological
birefringence, and constrains the length scale for any possible effect
to be at least several times larger than the length scale estimated by Nodland
and Ralston.  We thus find the original data to be completely consistent
with the more recently analyzed data \cite{Leahy97,WPC97}.

The data in question consist of four quantities for each of 160 radio galaxies
and was compiled by Carroll, Field, and Jackiw from previously published
observations \cite{CFJ90}.  The quantities are:  $\chi_i$, the 
intensity-averaged polarization angle for the radio emission from galaxy $i$,
with the effects of Faraday rotation removed;
$\psi_i$, the position angle for the galaxy describing the orientation
of its image on the sky; $\drxn_i$, the direction to the galaxy;
and $z_i$, the redshift of the galaxy.  Nodland and Ralston posited that
the polarization vector of light from a source at redshift $z$ along
direction $\drxn$ is rotated by an angle $\beta$ given by
\begin{equation}
\beta(z,\drxn) = {1 \over 2\lambda} \rho(z) \drxn\cdot\svec,\label{beta-def}
\end{equation}
where $\rho(z)$ is the luminosity distance to a source at redshift
$z$ in a flat universe in units of the Hubble distance, 
$\lambda$ is a 
length scale (also in units of the Hubble length) determining the magnitude
of the effect, and $\svec$ is a fiducial direction associated with
the effect.  They used a linear
correlation statistic to search for correlations between two quantities
derived from the data for each object:  a signed rotation angle, $\beta_i^\pm$,
measuring the rotation of polarization for galaxy $i$ \cite{betapm}, and  
$\rho(z_i) \drxn_i\cdot\svec$, the distance to the galaxy, projected
along a candidate fiducial direction $\svec$.  They determined the
significances of correlations in 410 candidate $\svec$ directions
using two different procedures involving simulated data based on the null
hypothesis of no correlation between $\chi_i$ and any other galaxy
parameters.  Our criticisms of their analysis fall into three categories.

First, their choice and use of statistics is ad hoc to a troubling
degree.  To be sure, there are no general criteria in frequentist
statistics guiding the choice of statistic in realistically complicated
problems, and in this sense the results of any study will be subjective.
But the Nodland and Ralston choices of the linear correlation
statistic $R$ using the signed angles $\beta_i^\pm$ creates apparent 
correlations when none exist (as they themselves acknowledged).
Their calibration using simulated data ideally should
account for this, but one might hope better choices are
available that do not create spurious effects by construction.  
In particular, since $\chi_i$ and $\psi_i$ are axial
variables (angles that take on values in $[0,\pi)$), the rotation 
from $\psi_i$ to $\chi_i$ is known only modulo $\pi$ and is thus
ambiguous.  Nodland and Ralston choose a priori to resolve the ambiguity in the
manner that most favors birefringence.  Also, their choice of 
the linear correlation statistic seems less than optimal considering
that one of quantities correlated is an axial variable, and in some
sense should be considered to ``wrap'' with a period of $\pi$.  Finally,
they use the correlation statistic both for hypothesis testing
and for parameter estimation.  There is an accessible
literature on the statistics of directional data that provides
more appropriate choices \cite{DData}, particularly for parameter
estimation, for which Nodland and Ralston use a specific model
that is amenable to analysis with standard estimation techniques.

Second, their choice of null hypothesis is inappropriate.  Others
have independently made this criticism \cite{EB97,CF97}.  
Their Monte Carlo simulations created data with $\chi_i$ and $\psi_i$ chosen
independently and randomly over $[0,\pi)$.  However, this is not
the only possible null hypothesis without birefringence, nor is it
a particularly interesting or realistic one.  One could detect birefringence 
only if $\chi_i$ and $\psi_i$ were correlated in the {\it absence} of
birefringence; one must presume that $\psi_i$ determines a preferred direction
for $\chi_i$, with cosmological birefringence causing a rotation away
from this.  It is only because a high degree of correlation is
both observed and expected between $\chi_i$ and $\psi_i$ that various
investigators have used this data to search for evidence of birefringence.
By their choice of null hypothesis Nodland and Ralston essentially
set up a ``straw man'' whose near-certain demise need not imply the existence
of birefringence.

The most appropriate null hypothesis is thus one that correlates
$\chi_i$ and $\psi_i$ with each other (but not with direction or redshift).
The existing literature offers a number of possibilities.
For example, several studies of the polarization
of light from radio galaxies explicitly point out that there appear
to be two populations of souces, a population with the polarization
nearly perpendicular to the galaxy orientation, and a population with
the polarization not very correlated with galaxy orientation, though
with a possible weak preference for parallel orientation.  
These facts, suggesting a reasonable and simple non-birefringent
null hypothesis, were clearly summarized by
Carroll et al.\ \cite{CFJ90}\ in their
presentation of the catalog analyzed by Nodland and Ralston.
Unfortunately, the manner in which Nodland and Ralston resolve
the modulo $\pi$ ambiguity in the polarization rotation maps the
data only to the first and third quadrants in the
$(\beta_i^\pm,\rho(z_i) \drxn_i\cdot\svec)$ plane, forcing their linear
correlation statistic to consider only lines through the origin,
i.e., to presume that the polarization vector should
be aligned with the galaxy major axis in the absence of birefringence
($\beta_i^\pm$ is presumed to vanish at low redshift, an observation
also made in \cite{CF97}).  Since the data are known
to prefer perpendicular alignment, a large correlation must result,
due not to a real correlation of rotation angle with redshift,
but rather to a poor choice of statistic and null hypothesis.
Even a real correlation between rotation angle and redshift need not
imply birefringence.
As Carroll et al.\ noted \cite{CFJ90}, the perpendicularly aligned
population consists of the more luminous galaxies, and is
thus visible to higher redshifts than the broader, possibly
parallel-aligned population.
Thus the typical alignment angle could vary with redshift simply
due to the selection effect that luminous galaxies are visible to
larger redshifts, so that perpendicularly aligned sources appear
preferentially at large redshift.  Any claim of evidence for birefringence
must take such reasonable ``null'' isotropic models into account.
If there is a significant correlation, assessment of such a
hypothesis could require explicit
inclusion of the galaxy luminosities in the analysis.

Finally, our third criticism of the Nodland-Ralston analysis is
that in their main Monte Carlo procedure they incorrectly 
assessed the significance with which
they rejected the null hypothesis.  They find the significances
associated with each of 410 candidate $\svec$ directions, and report
the smallest value as their main result (a probability for falsely
rejecting the null---the Type I error probability---of $\lesssim 10^{-3}$).
This value fails to account
for the fact that 410 directions were searched because the value
of $\svec$ is uncertain a priori.  Although the results
of the analyses for all 410 directions are unlikely to be independent,
the actual Type I error probability is certainly many times larger
than the value they report.  The second procedure they perform more
appropriately accounts for the size of the parameter space they
searched by performing a similar search for each simulated data set.
It is in fact the only reasonable procedure of the two they
use.  However, they apply it only to subsets of the data at
low and high redshift, finding a less significant
departure from the null hypothesis for the high redshift subset
and no significant departure from the null for the low redshift subset,
yet concluding these results
corroborate their earlier analysis (presumably because the data sets
are smaller so only smaller departures are expected).
The significance found by application of their second procedure
to the {\it entire} data set would be by far the most interesting quantity
they could calculate with their approach, but it remains curiously
absent from their analysis.

Many of these problems could be overcome with a careful frequentist
analysis of these data; some are addressed in the recent analysis of
Carroll and Field \cite{CF97}\ disputing the Nodland and Ralston results.
Here we instead pursue a Bayesian approach to the problem \cite{Bayes}.
It has the virtue of circumventing many of the problems cited
above in a largely ``automatic'' fashion.  We thus intend our study
not only to address the scientific
question of the existence of evidence for birefringence, but also to
offer a new tool for the study of polarization angle data that we
believe to be superior to existing tools in many respects.

Data enter a Bayesian analysis through the likelihood function---the
probability for the data given some hypothesis or model, $M$,
which in general may have some unknown parameters $\pars$.
We presume that, once a model is specified, the joint probability
for the data for all sources is simply the product of individual
probabilities for a single source's data, 
$p(\chi_i,\psi_i,\drxn_i,z_i\mid \pars, M)$.
We can factor this probability as follows:
\begin{equation}
p(\chi_i,\psi_i,\drxn_i,z_i\mid \pars, M) =
 p(\chi_i\mid \psi_i,\drxn_i,z_i, \pars, M) p(\psi_i,\drxn_i,z_i\mid \pars, M).
\label{pi-fac}
\end{equation}
If we were interested in comparing hypotheses that explicitly
modeled the distribution of source positions in space, or that 
sought to detect anisotropy in the distribution of source
directions or orientations,
the last factor would be important.  Instead, we here focus only
on isotropy evident in polarization angle rotation, as did
Nodland and Ralston.  Accordingly, we can ignore the last factor.

Our task thus becomes one of specifying models that allow us to
calculate the individual source likelihood functions, 
$\like_i(\pars) \equiv p(\chi_i\mid \psi_i,\drxn_i,z_i, \pars, M)$.
The full likelihood is just the product of all $\like_i(\pars)$ functions;
$\like(\pars) = \prod_i \like_i(\pars)$.
Note that we must specify a probability for the actual datum, $\chi_i$
(given $\psi_i$, $\drxn_i$, and $z_i$),
and not for a derived quantity such as the
signed rotation $\beta_i^\pm$ introduced by Nodland and Ralston in
an ad hoc attempt to resolve modulo $\pi$ ambiguities.  As will become
apparent, our approach lets the data determine which possible resolution
of the ambiguity is most probable for each object and each choice
of hypothesis.

We now must specify choices for $\like_i$.  
One choice is to take $p(\chi_i\mid
\psi_i,\drxn_i,z_i, \pars, M)$ to be a flat distribution, so that
$\like_i = 1/\pi$; this is the null hypothesis considered by Nodland
and Ralston.  We consider this uncorrelated
hypothesis ourselves, but we must specify more; no Bayesian calculation
can assess a ``null'' hypothesis without specific consideration of
one or more alternatives.  This may appear to be a serious drawback
of the Bayesian approach,
since tests like that used by Nodland and Ralston appear capable of
rejecting a null hypothesis without making subjective choices of
alternatives.  But this apparent advantage of the frequentist
hypothesis testing approach is largely illusory.  Subjectivity enters
the frequentist test in the choice of a test statistic.  Statisticians
have long appreciated that the choice of a particular
test statistic can often be considered to represent an implicit choice
of a particular class of alternative hypotheses, in the sense that a Bayesian
comparison of the null hypothesis to that class of alternatives results
in use of the same statistic used in the frequentist test (similar
conclusions apply in a maximum likelihood setting).  
We thus exchange subjectivity in the choice of statistic for subjectivity
in the choice of models.  This has the virtue of making the assumptions
underlying one's analysis explicit, which in turn can guide the
specification of further models that usefully generalize the analysis.

Jeffreys \cite{Jeffreys}\ showed that Bayesian
inference with conditional distributions that are Gaussian and
correlated along straight lines leads
to use of the linear correlation statistic.
Since we seek to mimic in a Bayesian setting the analysis Nodland and
Ralston attempted in a frequentist setting, their use of the
correlation statistic might suggest
that we introduce straight-line models and Gaussian errors. 
However, we are not free to introduce a priori
the signed rotation, $\beta_i^\pm$, they used in their correlation analysis;
any model we choose must specify a probability for the actually
observed ``raw'' data.  Also,
since $\chi_i$ is an axial quantity, we cannot use a Gaussian
distribution for it, but must instead use an axial distribution that
recognizes that $\chi_i$ is restricted to the interval $[0,\pi)$.
Since no correct Bayesian analysis of these data can lead precisely
to their choice of statistics, we instead seek to mimic
the spirit of their analysis, introducing linear models where
appropriate and using distributions that generalize the Gaussian
to angular variables.

Our models presume that
$\chi_i = (\psi_i + \theta_i) \mod \pi$, where $\theta_i$ is the 
actual angle of rotation of the polarization vector from the object
direction $\psi_i$, including a possible birefringent term.
The probability distribution for $\theta_i$ is a circular (not axial) 
distribution (i.e., its domain is $[0,2\pi)$) that can depend on
$\drxn_i$ and $z_i$.  From this assumption
it is straightforward to derive the axial probability distribution
for the observable $\chi_i$ from any hypothesized distribution
$f(\theta_i)$ for the rotation angle $\theta_i$:
\begin{equation}
\like_i(\pars)
 = f(\Delta_i) + f(\Delta_i + \pi),\label{like-gen}
\end{equation}
where
\begin{equation}
\Delta_i = \cases{\chi_i-\psi_i,&if $\chi_i \ge \psi_i$;\cr
    \pi-(\psi_i-\chi_i),&otherwise.\cr}\label{Delta-def}
\end{equation}
The quantity $\Delta_i$ is similar to $\beta_i^\pm$ of Nodland and Ralston.
But it was derived from an explicit model for the polarization rotation
rather than specified ad hoc,
and the likelihood factor for each galaxy explicitly accounts for
the two possible resolutions of the modulo $\pi$ ambiguity
associated with $\chi_i$.

To complete specification of a model, we must specify a
distribution for the rotation angle $\theta_i$.
To follow the spirit of the Nodland and Ralston analysis, we construct
candidate distributions using a well known circular generalization of
the Gaussian distribution, the Von Mises distribution \cite{DData}.
The Von Mises distribution
for an angle $\theta$ concentrated at angle $\theta_0$
with concentration parameter $\kappa$ is
\begin{equation}
f_{\rm VM}(\theta; \theta_0,\kappa) = 
{1 \over 2\pi I_0(\kappa)} e^{\kappa \cos(\theta-\theta_0)},\label{VM-def}
\end{equation}
where the normalization constant requires evaluation of a
modified Bessel function, $I_0(\kappa)$.  When $\kappa \gg 1$
this distribution becomes approximately Gaussian, with a mean
of $\theta_0$ and a variance of $1/\kappa$.

For our simplest nontrivial model, expressing a possible relationship between
$\chi_i$ and $\psi_i$ in the {\it absence} of birefringence,
we take $f(\theta_i)$ to be a Von Mises
distribution with unknown mean $\theta_m$ and concentration $\kappa$; both
parameters will be estimated using the data.  We denote
this model $M_1$.  Using equation~(\ref{like-gen}),
this corresponds to an object likelihood function
$\like_i(\theta_m,\kappa) = g(\Delta_i;\theta_m,\kappa)$, where
\begin{equation}
g(\theta; \theta_0,\kappa) = 
{\cosh\left[\kappa \cos(\phi-\phi_0)\right] \over 2\pi I_0(\kappa)}.\label{VM1}
\end{equation}
Contours of the resulting joint posterior distribution for
$\theta_m$ and $\kappa$ appear in Fig.~1; Table~1 lists the most probable
parameter values.  Here and throughout
this paper, posteriors are found by multiplying the likelihood
by prior distributions that are flat with respect to the plotted
axes, and normalized over the range plotted.  In this case,
the prior is flat over $\theta_m$ and flat over the {\it logarithm}
of $\kappa$ (the standard scale-invariant prior for an a priori
unknown nonzero scale parameter).

As is clear from the figure, the data strongly favor polarization
angles that are oriented {\it perpendicular} to the galaxy's
major axis, as one might have anticipated from visual inspection
of the data \cite{CFJ90}.  The distribution of angles is moderately 
concentrated, with the mode at $\kappa = 2.0$ (corresponding to
an angular half-width $\approx 40^\circ$).  Note that distributions
that preferentially align $\chi_i$ with $\psi_i$,
as was implicitly presumed in the Nodland and Ralston analysis, are
strongly ruled out.  It is also evident from this figure that
the null hypothesis of no relationship between $\chi_i$ and
$\psi_i$ (corresponding to $\kappa = 0$) is strongly ruled out.
In the Bayesian framework we can quantify this by calculating
the posterior odds in favor of the Von Mises model $M_1$ over the null
hypothesis (i.e., the ratio of their posterior probabilities).
The posterior odds is the product of a subjective prior odds
factor, and a data-dependent term called the Bayes factor.
The Bayes factor is the ratio of the global likelihoods for the competing 
models, where the global likelihood for a model with free parameters is 
found by averaging the likelihood for its parameters (weighted by the prior 
distributions for the parameters).
This is in contrast to frequentist hypothesis testing methods
which instead extremize over parameter space (e.g., maximum
likelihood or minimum $\chi^2$).  The averaging
required by the Bayesian approach implements an automatic
``Ockham's razor'' that penalizes models for the size of
their parameter space.  As a result, even though a complicated
model may fit the data better than a simpler model in the sense of having 
a larger maximum likelihood, it can be {\it less} probable than 
the simpler model if the likelihood is not increased enough to account
for the larger size of the parameter space of the more complicated
model.  It is thus customary in Bayesian literature to report
Bayes factors for model comparison, essentially presuming the
prior odds to be unity (since model complexity is already
incorporated into the Bayes factor).  If one judges
certain models to be significantly less probable than others
a priori (on grounds other than their complexity), one can
multiply the reported Bayes factor to account for this.

For the null hypothesis, $\like_i = 1/\pi$, and there are no
free parameters; we denote this model as $M_0$.  
The likelihood for this model is
thus simply $1/\pi^N$, where for the full dataset $N=160$.
For the Von Mises model, $M_1$, one must average
the numerically calculated likelihood over $\theta_m$ and $\kappa$.
The resulting Bayes factor is $1.9\times 10^4$ in favor of $M_1$ 
over $M_0$, conclusively favoring a correlation between $\chi_i$
and $\psi_i$ over the null hypothesis of no correlation.

We now introduce cosmological birefringence into the Von Mises
model by setting
$\like_i(\theta_m,\kappa) = g(\Delta_i;\theta_m+\beta_i,\kappa)$, 
where the expected angle of rotation for object $i$ is shifted
by $\beta_i = \beta(z_i,\drxn_i)$ (c.f., eqn.~(\ref{beta-def})),
as posited by Nodland and
Ralston.  This adds three parameters to the model, the
dimensionless length scale $\lambda$, and the unknown
fiducial direction $\svec$ (specified by right ascension $\alpha$
and declination $\delta$).  We denote this model as $M'_1$.  
We use flat priors for $\alpha$ and
$\sin(\delta)$ and a flat prior for $\log(\lambda)$ with
$0.01, < \lambda < 10$ (corresponding to distance scales appropriate for
detecting a cosmological effect; i.e., at least a reasonable
fraction of a Hubble length, up to the length scale of the
observable universe).  The solid curve in
Fig.~2 shows the resulting marginal
posterior distribution for $\lambda$.  This marginal distribution 
$p(\lambda\mid D,M'_1)$
is found by numerically integrating the full joint posterior distribution
$p(\lambda,\alpha,\delta,\theta_m,\kappa\mid D,M'_1)$
over the remaining four parameters 
$\alpha,\delta,\theta_m,\kappa$, and thus accounts
for parameter uncertainties and correlations between all parameters
\cite{adapt}.
For $\lambda > 1$, the marginal posterior is very nearly
flat, i.e., it resembles the prior.  In other words, in
this region the data have taught us nothing---they are 
uninformative about birefringence on these length scales,
as one might have guessed a priori since there are few
galaxies in the data set with $z>1$.  But in the
$\lambda < 1$ region, the data are very informative.
The 95.4\% (``$2\sigma$'') highest posterior density
credible region (CR) begins at $\lambda = 0.43$ and
extends to the upper limit of the prior range.  Birefringent
models with $\lambda \lesssim 0.4$ are thus conclusively
ruled out.  The best-fit model of Nodland and Ralston
had $\lambda \approx 0.1$.

The Bayes factor comparing this model to its simpler isotropic
counterpart is 0.46.  Though slightly favoring
the isotropic model, this Bayes factor indicates that the
data do not conclusively discriminate between birefringent
and isotropic Von Mises models.  However, the allowable
birefringent models are those about which the data are
uninformative (those with $\lambda \gtrsim 1$).  The models
with small $\lambda$ favored by Nodland and Ralston are conclusively
rejected.

Finally, we note that the best-fit values of the parameters for
the underlying polarization angle distribution are little changed from
their best-fit values in the isotropic model (see Table~1).
Thus the preference for perpendicular alignment is not a consequence
of ignoring possible birefringence.
The best-fit fiducial direction is at $\alpha=149^\circ$
and $\delta=11^\circ$, very different from the Nodland and
Ralston best-fit direction of $\alpha=315^\circ$ and
$\delta = -10^\circ$.  However, the posterior varies
only weakly with $\svec$, so the fiducial direction is
highly uncertain and can lie almost anywhere on the sky.
This is consistent with there being negligible evidence
for birefringence in these data.

We also considered models that are more complicated than those implicit
in the Nodland and Ralston analysis.  These models
were motivated by the possibility \cite{CFJ90} that there are two
populations of sources, one with polarization preferentially perpendicular
to the source orientation (as was infered in the model
analyzed above), and another with polarization only weakly
aligned, with a possible preference for parallel alignment.
Accordingly, we considered
models that superposed parallel and perpendicular populations by taking
\begin{equation}
\like_i(\theta_m,\kappa) = f g(\Delta_i;0,\kappa_\|)
+ (1-f) g(\Delta_i;\pi/2,\kappa_\bot),\label{VM2}
\end{equation}
where $f$ is the fraction of sources in the parallel population,
and $\kappa_\|$ and $\kappa_\bot$ are the
widths of the parallel and perpendicular populations.
We studied this model and two simpler cases of it: a model
combining the parallel population with a population with
a flat $\Delta_i$ distribution (i.e., fixing $\kappa_\bot=0$),
and a model combining the perpendicular population with a
flat $\Delta_i$ distribution (i.e., fixing $\kappa_\|=0$).  Of these
three classes of models, only the latter was favored by the
data over the single population model discussed above (regardless
of whether birefringence was added or not).  We here discuss
results only for this model, which we denote $M_2$.

Model $M_2$ has two parameters, $f$ and $\kappa_\bot$.
Fig.~3 shows contours of the joint posterior for these parameters.
Table~1 lists the most probable values.
There is a clear mode corresponding to roughly half of the
sources coming from a flat population and the remainder coming
from a perpendicular population that is moderately
concentrated with $\kappa_\bot \approx 4$ (corresponding to
a half-width of $\approx 30^\circ$).  But the posterior
has a tail extending to low $f$ with $f=0$ lying within the
95.4\% CR, suggesting that the data do not decisively prefer
this model to a single population model.  The Bayes factor
comparing this model to the single population model discussed
above is 54, indicating significant but not conclusive
evidence for the addition of a flat population.

Next we considered a birefringent version of this model,
$M'_2$, by taking
\begin{equation}
\like_i(\theta_m,\kappa) = {f \over \pi}
+ (1-f) g(\Delta_i+\beta_i;\pi/2,\kappa_\bot).\label{SVM2}
\end{equation}
This model has five parameters:  $f$, $\kappa_\bot$, $\lambda$,
and $\svec = \{\alpha,\delta\}$.  The Bayes factor comparing
this model to its counterpart without birefringence is 
$\approx 0.44$, again indicating that though the data prefer
models without birefringence, they do not conclusively rule
it out.  However, as with single component models, small
values of $\lambda$ are ruled out, as is apparent from the
marginal distribution shown as the dashed curve in Fig.~2
(the lower limit of the 95.4\% CR is at $\lambda = 0.52$).
Again, values of the order of that found by Nodland and
Ralston are excluded with high probability.

Following Nodland and Ralston, we repeated the analysis using
only the subset of 71 sources with $z \ge 0.3$, the redshift
where they claim the birefringence effect ``turns on.''
Table~2 lists the resulting parameter estimates and Bayes factors.
None of the conclusions found in our analysis of all 160 sources
was changed.  Bayes factors again indicate that isotropic models
are prefered to their birefringent counterparts, though
not decisively so.  The credible regions for $\lambda$ again
decisively rule out the Nodland and Ralston value; the lower
limit of the 95.4\% CR is at $\lambda \approx 0.47$ for the
single component model $M'_1$, and at $\lambda \approx 0.62$ for
model $M'_2$ combining a perpendicular population with a flat
population.  Note that the estimated fraction in the flat
polulation is significantly smaller than was the case
in the analysis of all 160 galaxies.  This may support the
observation of Carroll, Field, and Jackiw \cite{CFJ90}\ that
the sources with polarization perpendicular to the galaxy
orientation tend to have higher luminosities (and thus dominate
the sample at high redshifts).

We emphasize that these results have been obtained using the
same data analyzed by Nodland and Ralston, with the same
explicit model for birefringence, and with models for the
underlying relationship between $\chi_i$ and $\psi_i$ as close
as possible to what was implicitly presumed in their analysis.
Our findings reveal
the data in question to be completely consistent with more
recently analyzed data that constrain any cosmological birefringence
to be much smaller in magnitude than that claimed by Nodland
and Ralston \cite{Leahy97,WPC97}.  The primary difference
between the Nodland-Ralston study and ours is methodological.  
Our methodology improves
on the frequentist one adopted by Nodland and Ralston in all
three categories mentioned above.  Our choice of statistics is
dictated by our choice of models via the rules of probability
theory and rigorously accounts for the axial nature of the
observed quantities, and in particular for modulo $\pi$
ambiguities.  We considered a variety of hypotheses with and
without birefringence, and thus have not fallen prey to
claiming detection of birefringence by rejecting an unnecessarily
simple ``straw man'' null hypothesis.  And finally, the
Bayesian methodology fully and automatically accounts for
the sizes of the parameter spaces of our models and for all
correlations between inferred parameters, both in the averaging
process underlying calculation of Bayes factors for comparing
models, and in the integrations required in calculating marginal
distributions for subsets of parameters of particular interest
(e.g., $\lambda$).
Quite apart from the issue of the existence of birefringence,
we believe these qualities strongly recommend Bayesian methods
for the analysis of polarization angle data.  They can be
straightforwardly generalized to address questions beyond those
dealt with in this brief paper.  For example, uncertainties
in the measured quantities can be straightforwardly accounted
for by marginalizing.  Also, more sophisticated models can
be studied by
including data on the polarization strengths and luminosities
of sources in order to determine whether alignment is correlated
with these quantities, and whether population parameters change
with redshift in a manner that can be attributed to source
evolution, or in a manner that instead demands the existence of
cosmological anisotropy.

We thank Sean Carroll for providing a machine-readable catalog of
the polarization data published in \cite{CFJ90}, and Borge Nodland
for information about typographical errors in the catalog.
This work was supported in part by NASA grants NAG 5-3427 and 
NAG 5-3097 and by NSF grants AST 91-19475, AST 93-15375,
and PHY-9408378.


\newpage

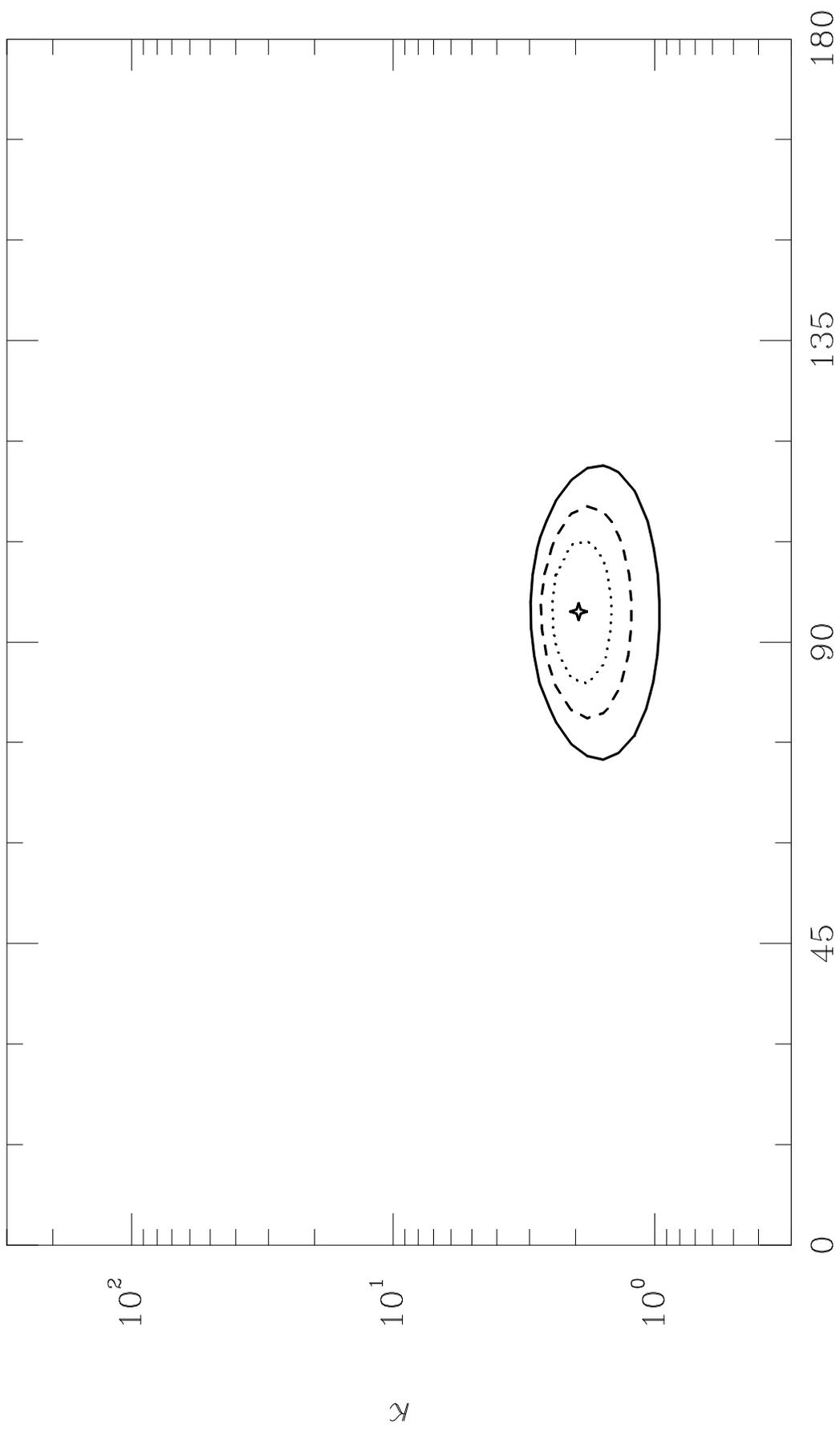
\begin{figure}
\caption{Joint posterior distribution for $\theta_m$ and $\log_{10}(\kappa)$
in the single component model $M_1$,
with no birefringence.  Contours indicate the boundaries of
68.3\% (dotted), 95.4\% (dashed), and 99.73\% (solid) credible regions.
Cross indicates the mode.}
\label{fig-M1}
\end{figure}

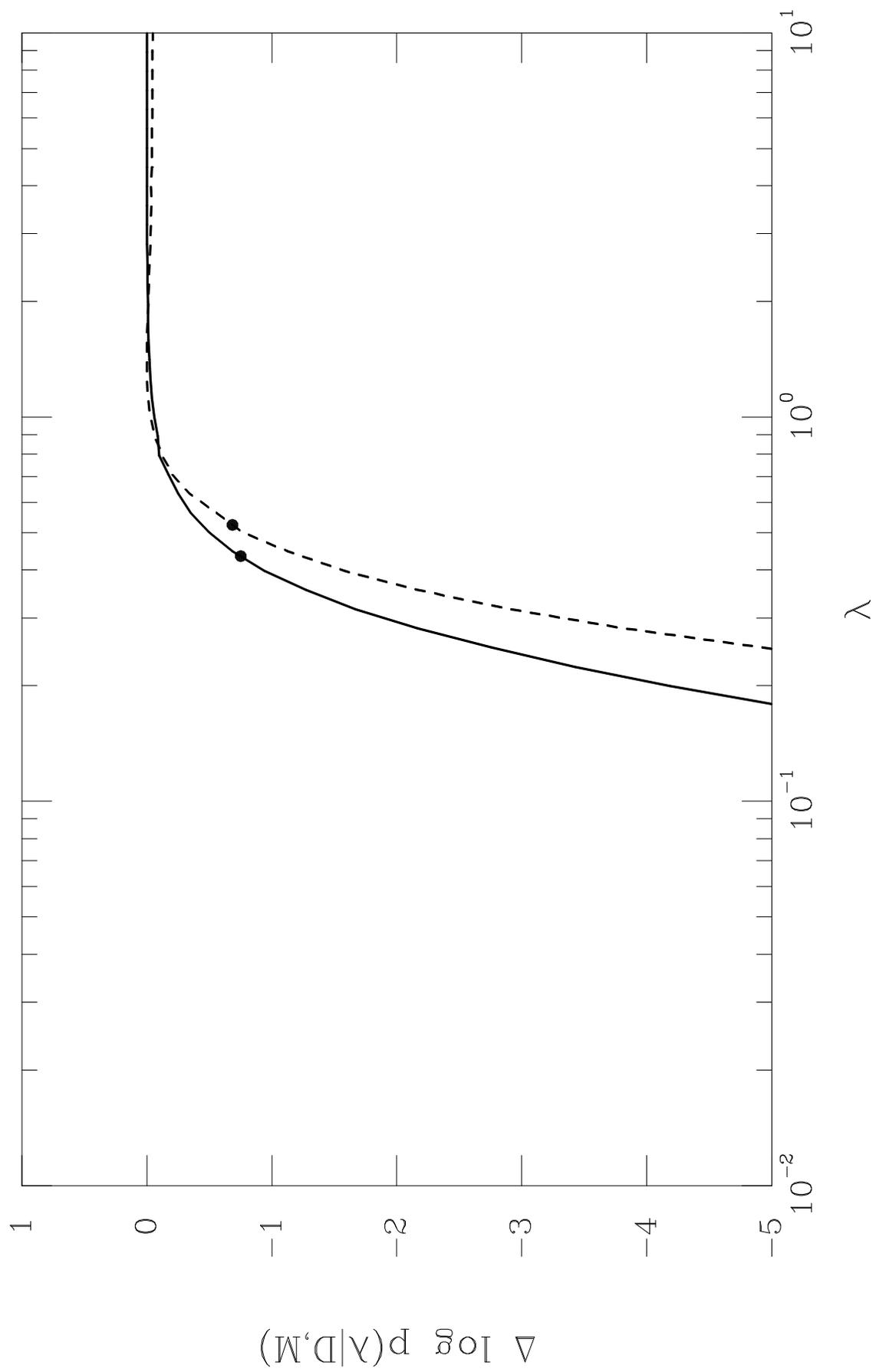
\begin{figure}
\caption{Logarithm of marginal posterior distributions for 
the birefringence lengthscale $\lambda$ (in units of the Hubble distance),
with arbitrary normalization.  Dots mark the lower limits
of 95.4\% credible regions.  The solid curve is for the single component 
birefringent model $M'_1$; the dashed curve is for the two component
birefringent model $M'_2$.}
\label{fig-lambda}
\end{figure}

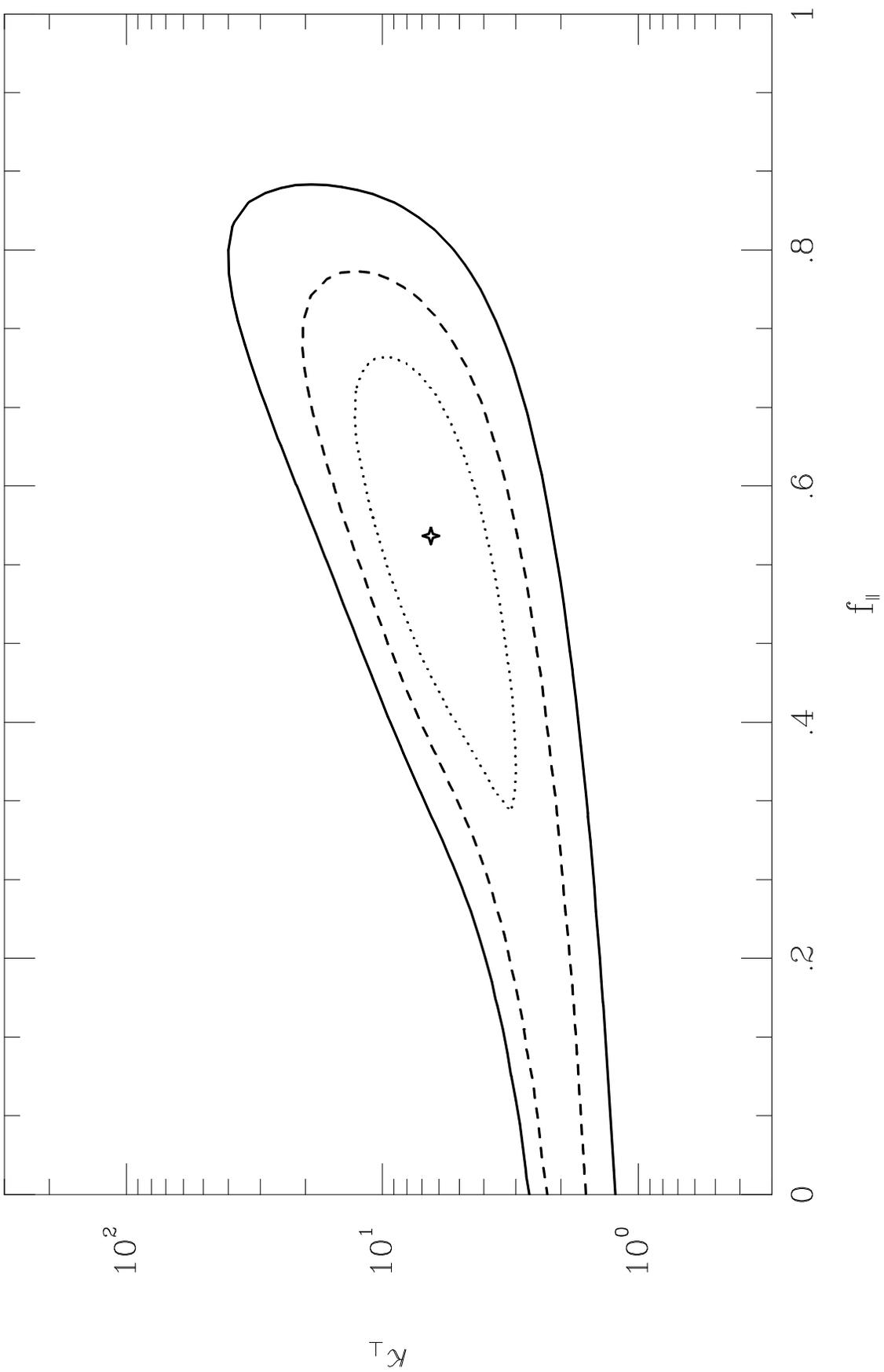
\begin{figure}
\caption{Joint posterior distribution for $f$ and $\log_{10}(\kappa_\bot)$ in
the two component model $M_2$,
with no birefringence.  Contours indicate the boundaries of
68.3\% (dotted), 95.4\% (dashed), and 99.73\% (solid) credible regions.
Cross indicates the mode.}
\label{fig-M2}
\end{figure}

\newpage

\narrowtext
\begin{table}
\caption{Most probable parameter values and Bayes factors;
all data (160 galaxies)\label{table1}}
\begin{tabular}{lcc}
Quantity &  Isotropic & Birefringent  \\
\tableline
\multicolumn{3}{c}{$M_1, M'_1$: Single Von Mises}\\
$\theta_m$ $(^\circ)$ & 94.6 & 91.3  \\
$\log_{10}\kappa$  & 0.29 & 0.31  \\
$B$\tablenotemark[1] & \multicolumn{1}{c}{$1.9 \times 10^{4}$} &  %
         \multicolumn{1}{c}{$8.5 \times 10^{3}$}  \\
\noalign{\bigskip}
$\log_{10}\lambda$ & --  & -0.15  \\
$\alpha$ $(^\circ)$ & -- & 149 \\
$\sin \delta$& -- & 0.19 \\
\tableline
%
\multicolumn{3}{c}{$M_2, M'_2$: Perpendicular Von Mises $+$ Constant}\\
$f$ & 0.56 & 0.55 \\
$\log_{10}\kappa_\bot$  & 0.81 & 0.84  \\
$B$\tablenotemark[1] & \multicolumn{1}{c}{$1.0 \times 10^{6}$} &  %
         \multicolumn{1}{c}{$4.5 \times 10^{5}$}  \\
\noalign{\bigskip}
$\log_{10}\lambda$ & --  & -0.13  \\
$\alpha$ $(^\circ)$ & -- & 146 \\
$\sin \delta$& -- & 0.23 \\
\end{tabular}
\tablenotetext[1]{Bayes factors compare the models to model $M_0$
positing a flat polarization angle distribution, uncorrelated with
position angle.}
\end{table}

\narrowtext
\begin{table}
\caption{Most probable parameter values and Bayes factors;
data selected with $z\ge 0.3$ (71 galaxies)\label{table2}}
\begin{tabular}{lcc}
Quantity &  Isotropic & Birefringent  \\
\tableline
\multicolumn{3}{c}{$M_1, M'_1$: Single Von Mises}\\
$\theta_m$ $(^\circ)$ & 93.7 & 85.8  \\
$\log_{10}\kappa$  & 0.56 & 0601  \\
$B$\tablenotemark[1] & \multicolumn{1}{c}{$2.4 \times 10^{7}$} &  %
         \multicolumn{1}{c}{$1.3 \times 10^{7}$}  \\
\noalign{\bigskip}
$\log_{10}\lambda$ & --  & -0.25  \\
$\alpha$ $(^\circ)$ & -- & 155 \\
$\sin \delta$& -- & 0.64 \\
\tableline
%
\multicolumn{3}{c}{$M_2, M'_2$: Perpendicular Von Mises $+$ Constant}\\
$f$ & 0.17 & 0.051 \\
$\log_{10}\kappa_\bot$  & 0.71 & 0.63  \\
$B$\tablenotemark[1] & \multicolumn{1}{c}{$2.2 \times 10^{8}$} &  %
         \multicolumn{1}{c}{$9.6 \times 10^{7}$}  \\
\noalign{\bigskip}
$\log_{10}\lambda$ & --  & -0.12  \\
$\alpha$ $(^\circ)$ & -- & 155 \\
$\sin \delta$& -- & 0.38 \\
\end{tabular}
\tablenotetext[1]{Bayes factors compare the models to model $M_0$
positing a flat polarization angle distribution, uncorrelated with
position angle.}
\end{table}


\begin{references}
\bibitem{betapm}
More precisely, Nodland and Ralston took 
$\beta_i^\pm = \chi_i - 
\psi_i + n \pi$ where $n$ is an integer such that $0 \le \beta_i^+ \le
\pi$ and $-\pi \le \beta_i^- \le 0$; for each galaxy they chose
between $\beta_i^+$ and $\beta_i^-$ according to the sign
of $\drxn_i\cdot\svec$ for the candidate $\svec$ under consideration.
\end{references}
\end{document}